# Cross-Domain Influences on Creative Processes and Products

**Victoria Scotney (v.scotney@alumni.ubc.ca),**

**Sarah Weissmeyer (sarah.weissmeyer@alumni.ubc.ca),**

**and Liane Gabora (liane.gabora@ubc.ca)**

Department of Psychology, University of British Columbia
Okanagan Campus, Fipke Centre for Innovative Research, 3247 University Way
Kelowna BC, Canada V1V 1V7

## Abstract

According to the honing theory of creativity, the iterative process culminating in a creative work is made possible by the self-organizing nature of a conceptual network, or worldview, and its innate holistic tendency to minimize inconsistency. As such, the creative process is not limited to the problem domain, and influences on creativity from domains other than that of the final product are predicted to be widespread. We conducted a study in which participants with varying levels of creative experience listed their creative outputs, as well as influences (sources of inspiration) on these outputs. Of the 758 creative influences, 13% were within-domain narrow, 13% within-domain broad, 67% cross-domain, and 6% unclear. These findings support the hypothesis that to trace the inspirational sources or 'conceptual parents' of a creative output, and thus track its cultural lineage, one must look beyond the problem domain to the creators' self-organizing, inconsistency-minimizing worldview at large.

**Keywords:** art; creativity; cross-domain influence; domain-general; domain-specific; innovation; inspiration; music

## Introduction

Creativity is thought to involve the restructuring of information in a creative domain, sometimes referred to as the *problem domain* (Runco, 2014), or simply, the domain. *Domain-specific theories* of creativity emphasize the non-transferability of expertise from one creative domain to another (Baer, 2015). They appear to be supported by findings that creative individuals are rarely creative in more than a few domains, i.e., someone known for their creativity in physics is rarely also known for their creativity as a dancer (Baer, 2012; Kaufman & Baer, 2004a). Support also comes from studies in which individuals created products for different domains, such as poems and mathematical equations, which found a low correlation in the ratings of individuals' creativity across domains (Baer, 1991).

*Domain-general theories* emphasize the generalizability of creative thinking across different domains (Hong & Milgram, 2010). The domain general view is supported by personality studies, which suggest that there is something to the notion of a creative personality type (Batey & Furnham, 2006; Eysenck, 1993; Feist, 1998; Martindale & Daily, 1996), and by evidence that when people express themselves in different creative domains these outputs bear a recognizable style or 'voice' (Gabora, O'Connor, & Ranjan, 2012; Ranjan, 2014). Such findings suggest that the creative mind seeks to explore and express its distinctive structure and dynamics using whatever means available.

Currently, many scholars espouse a less dichotomous view of creativity that incorporates both domain-specific and domain-general elements (Gabora, 2017; Plucker & Beghetto, 2004; Kaufman & Baer, 2004c). Nevertheless, we believe that the domain-generality of creativity is still under-appreciated due to emphasis on the final *product* or output of the creative process. Even if creative individuals tend to express themselves in one domain this does not necessarily mean that prior phases of their creative process are domain-specific. This paper describes a study designed to test the hypothesis that cross-domain influences play a normal and natural role in creativity and constitute a ubiquitous part of the creative process, and that the prominence of cross-domain influences on creativity is not exclusive to expert creators.

## Honing Theory

*Central Aim and Core Concepts*. While the central aim of most theories of creativity is to account for the existence of creative products, the *honing theory of creativity* (HT) arose to account for the cumulative, open-ended nature of cultural evolution. HT grew out of the view that humans possess two levels of complex, adaptive, self-organizing, evolving structure: an organismic level, and a psychological level (Barton, 1994; Combs, 1996; Freeman, 1991; Gabora, 1998, 2017; Pribram, 1994; Varela, Thompson, & Rosch, 1991). We refer to this psychological level as a *worldview:* an individual's unique dynamic web of understanding that provides a way of both *seeing* the world and *being in* the world, i.e., a mind as it is experienced from the inside. In short, HT posits that the worldview is the hub of a second evolutionary process—cultural evolution—that rides piggyback on the first—biological evolution—and that creative thinking fuels this second evolutionary process (Gabora, 1998, 2004, 2008, 2013, 2017).

Honing an idea entails reiteratively looking at it from the different angles proffered by one's particular worldview, 'putting ones' own spin on it', and make sense of it in one's own terms, followed by expressing it outwardly (Gabora, 2017). Honing may involve the restructuring of representations by re-encoding the problem such that new elements are perceived to be relevant, or relaxing goal constraints (Weisberg, 1995), and self-organized criticality, wherein small perturbations can have large effects (Gabora, 1998). As the creator's understanding of the task shifts, the creative idea may find a form that fits better with the worldview as a whole, such that the worldview achieves a more coherent state, as formalized by the notion of *conceptual closure* (Gabora, 1998; Gabora & Steel, 2017). Creative acts and products render such cognitive transformation culturally transmissible. Thus, it is suggested that what evolves through culture is not creative contributions but worldviews, and cultural contributions give hints about the worldviews that generate them.

*Predictions of HT Concerning Cross-Domain Influences.* HT posits that creative output reflects the idiosyncratic, transformative process by which a worldview restructures itself in response to perturbations such as the detection of threats, inconsistencies, ambiguity, or potentiality. Such perturbations cause *arousal-provoking uncertainty* (Gabora, 2017), which Hirsh, Mar, and Peterson (2012) refer to as *psychological entropy*, and which sets self-organized iterative honing into motion. HT posits that since the contents of the mind collectively self-organize, none are *a priori* excluded from the creative task, and it is possible for the domain-specific aspects of an idea to be stripped away such that it is amenable to re-expression in another form.

Because a worldview can continuously renew its overall structure, there are no limits on the possible influences or 'conceptual parents' of a creative work such as a song or journal article. For example, consider the situation in which a book inspires a movie, which inspires an invention. To see the thread of continuity across this 'line of descent' it is necessary to consider how their creators navigate through webs of beliefs, attitudes, procedural and declarative knowledge, and habitual patterns of thought and action that emerge through the interaction between personality and experience. In short, HT predicts that cross-domain influences play a role in the creative processes that fuel the self-organized transformation of worldviews, and that this in turn is the driving force of cultural evolution.

## Previous Research on Creative Influences

There have been efforts to corroborate anecdotal reports of creative influences (see Feinstein, 2006) with machine learning techniques designed to resolve lines of influence (Saleh, Abe, & Elgammal, 2014). However, these techniques are not yet able to discern cross-domain influences, wherein a creator in one domain (e.g., artist) is influenced by another domain (e.g., music).

Chan, Dow, and Schunn (2015) examined the conceptual distance of inspirational sources on the quality of design ideas, and found that conceptually closer sources (which were defined as sharing a 'topic' of closely associated words) were associated with higher quality solutions. However, as the authors note, this finding is inconsistent with the relatively robust finding that problem solvers from outside a given problem domain often produce the most creative solutions (Franke, Poetz, & Schreier, 2014; Hargadon & Sutton, 1997; Jeppesen & Lakhani, 2010). Moreover, since the designers were not asked to provide all elements that inspired their work, the scope of study was limited to the kinds of influences that one might logically expect to have a direct bearing on the result. Thus, for example, they listed things like previously generated solutions or design ideas, but not things like a particular piece of music, or 'a conversation with a friend'.

Studies of cross-domain inspiration showed that it is possible to re-interpret a creative work in one medium into another medium (Ranjan, Gabora, & O'Connor, 2013; Ranjan, 2014). When painters were instructed to paint what a particular piece of music would 'look like' if it were a painting, naïve participants were able to correctly identify at significantly above chance which piece of music inspired which painting. Although the medium of expression was different, something of its essence remained sufficiently intact for people to detect a resemblance between the new creative output and its inspirational source. This suggested that, at their core, creative ideas are less domain-dependent than is widely assumed. The study supported our intuitive conviction that when the creative *output* is not a blend but lies squarely in one domain, the *creative process giving rise to it* may be rooted in different domains. However, due to the artificial nature of this study it did not provide evidence that cross-domain plays a role in real-world creative endeavors.

In a precedent to the current study, 66 individuals with demonstrable accomplishments in a fine arts domain (e.g., music, painting, or fiction writing) were asked to list, for each of their most significant creative works, all influences on the creative process of generating these works (Gabora & Carbert, 2015). Of the 65 creative influences provided by the 66 participants, 47% were cross-domain influences (e.g., a painting influenced by music), 27% were narrow within-domain (e.g., a painting influenced by another painting), 8% broad within-domain (e.g., a painting influenced by sculpture), and 18% unclassifiable. Thus, the cross-domain influences were more widespread than within-domain influences, even when broad within-domain influences (e.g., technology influenced by music) as well as narrow within-domain influences (e.g., music influenced by other music) were taken into account. This result was surprising, for we had just been looking to see if cross-domain influence exists at all; we were not expecting it to predominate.

A limitation of this previous study was that the sample size was small, and because it included only expert level creative individuals it did not enable us to make conclusions about creative processes beyond this exclusive group. The goal of the current study was to replicate the above study on a participant population that is not characterized by expertise in a creative domain.

## Method

### Participants

The participants were 463 undergraduate students (114 males, 347 females, and two who selected 'no or different gender') from the University of British Columbia. They were recruited through SONA, an online participant pool approved by the UBC Research Ethics Board.

**Procedure**

The SONA website provided a link to an online questionnaire hosted by FluidSurvey. After formally consenting to the study, participants were asked to provide their age, gender, and occupation. The questionnaire then asked the respondents to answer the following questions:

1. What is the general category for the creative work for which you are most known (e.g., art, music, drama, science)?
2. What is the subcategory for the creative work for which you are most known (e.g., painting, piano composition, biochemistry)?
3. Please describe your creative outputs.
4. Please describe as best you can your creative process.
5. Describe all elements that have inspired your work (natural or artificial, or it may be a particular event or situation, or something not in the concrete environment, that is, something abstract that you have been thinking about), and with each item, if possible, put as much identifying information as you can about the item it inspired (e.g., my Sunlight Sonata in B Flat composed in 2012 was inspired by going skiing in the alps with my sister who had just recovered from pneumonia). Do this for as many of your creative works as you can, itemizing them as (a), (b), (c), and so forth. Provide as much detail as possible.

The first three questions were used to determine the domains of creative outputs for each participant. Question five was used to assess the domain of the inspirational source associated with each creative output. Question four was not used in this study, as it did not provide additional clarification regarding either creative outputs or their inspirational sources. The questions were worded in an open-ended way so as not to constrain participants' answers in any way. The study was designed as an online survey, rather than a personal interview, so that we could maximize the time we had to collect as much data as possible.

**Analysis**

Of the 463 participants, 111 were excluded because they left one or more of questions one, two, three, or five blank. (They were able to receive partial course credit whether or not they completed the questionnaire.) An additional 90 questionnaires were excluded because the participants did not provide classifiable answers. Some of the questionnaires were unclassifiable because participants either provided a description of their creative process, or answered with what motivated a creative output, rather than providing an inspiration; e.g., an answer was excluded if the only reason for engaging in the creative activity was that someone else, such as a parent, obliged them to participate. The second category of unclassifiable questionnaires were those in which no creative outputs were provided in conjunction with a creative influence. The last category of unclassifiable questionnaires were those for which both raters independently found them incomprehensible and impossible to evaluate. This left 262 completed questionnaires for analysis, and these questionnaires provided a total of 758 influences.

By comparing the answers to questions one and two with those for question five, the provided influences were organized into four categories: within-domain narrow, within-domain broad, cross-domain, and uncertain. A response was classified as *within-domain narrow* (WN) if the domain of inspiration and the domain of the creative output fell within the same subcategory, e.g., a painting inspired by another painting. An example of a *within-domain broad* (WB) influence is a photograph inspired by a painting. A photograph and a painting belong to the same domain of visual art, but not to the same subdomain. *Cross-domain* (C) influences were those for which the domain of the influence was different from the domain of creative expression. For example, a song (domain: music) that inspired a software program (domain: technology) was rated as a cross-domain influence. An answer was categorized as *unclear* (U) when insufficient information was provided to categorize the influence as WN, WB, or C.

Two raters independently categorized these responses as either WN, WB, C, or U. Inter-rater reliability between the raters, calculated using both Cohen's kappa ($\kappa = .77$) and Krippendorff's alpha ($\alpha = .87$) was well beyond what would be expected by chance (see Davey, Gugiu & Coryn, 2010). Where there were any inconsistencies between the ratings of the two raters, they reached a final decision through discussion.

**Results**

The total number of influences in each domain category, as well as the upper and lower limits for the 95% confidence interval for each domain category, are provided in Table 1.

Table 1: Frequency and Percentage of Domain Influences on Creative Output and Lower Limit and Upper Limit for a 95% Confidence Interval.

| Creative Influence | Frequency | Lower Limit | Upper Limit | % of total |
|---|---|---|---|---|
| Within-Domain Narrow (WN) | 101 | 83.60 | 121.28 | 13.32% |
| Within-Domain Broad (WB) | 101 | 83.60 | 121.28 | 13.32% |
| Cross-Domain (C) | 508 | 481.53 | 533.15 | 67.02% |
| Unclear (U) | 48 | 36.02 | 63.44 | 6.33% |

The participants' creative outputs came from a variety of domains, including drawing, architecture, photography, scientific experiments, song writing, furniture design, biochemistry, and athletic performance. They also gave a wide variety of inspirational sources, ranging from people in their lives such as family, friends, and strangers, to African safaris and *The Book of Kells*. Each participant listed an average of 2.89 creative influences ($SD = 1.92$, $Mdn = 3$, range: 1-15).

Of the 758 influences provided, 101 (13.32%) were WN, 101 (13.32%) were WB, 508 (67.02%) were C, and 48 (6.33%) were U. Thus, cross-domain influences constituted more than double the number of within-domain and unclear influences, even when broad as well as narrow within-domain influences were included.

Since the participants were encouraged to list all the influences they could think of that inspired their creative outputs, there was variation in the number of influences that each participant provided. A possible explanation for the low frequency of WN influences could be that there is only one NW influence that can match a given creative output, but there is a larger number of WB influences, and a potentially infinite number of C influences. Thus, once someone has provided an influence that matches the domain of the creative output, any subsequent influences they provide will necessarily be either cross-domain or within-domain broad. To explore the possibility that our results give more weight to participants who gave multiple influences, we re-analyzed the data according to the number of influences provided by each participant, as shown in Table 2.

Table 2: Percentage and Frequency (in brackets) of Within-Domain Narrow (WN), Within-Domain Broad (WB), Cross-Domain (C) and Unclear (U) Influences by Total Number of Influences (N-Infl) Provided by Participant. N-Partic is the number of participants who provided that number of influences.

| N-Infl | N-Partic | WN | WB | C | U |
|---|---|---|---|---|---|
| 1 | 68 | 26.47% (18) | 7.35% (5) | 57.35% (39) | 8.82% (6) |
| 2 | 59 | 14.41% (17) | 15.25% (18) | 66.10% (78) | 4.24% (5) |
| 3 | 57 | 10.53% (18) | 13.45% (23) | 66.08% (113) | 9.94% (17) |
| 4 | 40 | 13.12% (21) | 14.38% (23) | 66.86% (107) | 5.62% (9) |
| 5 | 19 | 8.89% (8) | 13.33% (12) | 75.56% (68) | 2.22% (2) |
| 6 or more | 19 | 12.58% (19) | 13.25% (20) | 68.21% (103) | 5.96% (9) |

This re-analysis shows that there may be some merit to this explanation, for participants who gave only one influence did indeed have a higher frequency of WN influences than any other group (26.47% versus between 8.89 – 14.41%) and a lower frequency of C influences (57.35% versus between 66.08 – 75.56%). However, this does not alter the overall pattern of the findings that even when only one influence was given, that influence was more often than not (i.e., 57.35% of the time) a cross-domain influence.

## Discussion

The results of this study concur with previous finding (Gabora & Carbert, 2015) that the majority of creative outputs were inspired by cross-domain influences. However, the current results show that this is not just the case for individuals with proven success or expertise in a creative domain; it holds true for non-experts as well. This result has implications for our understanding of the creative process, because it demonstrates that it is substantially less domain-specific than it is widely presumed to be. Even if individuals primarily express their creativity in a single domain, they are often employing cross-domain thinking when they create. Although domain-specific knowledge may ensure that tools of the trade are appropriately applied, and one's creative works may consistently be in one particular domain, the sources that initially triggered these creative processes may be diverse in nature. To the extent that this is the case, creativity may involve synthesizing information from different arenas of one's life.

### Implications for a Theoretical Framework for Creativity

The finding that cross-domain influences are widespread is consistent with HT, according to which it is not just one's conception of the creative task (or 'problem domain') but one's worldview as an integrated whole that transforms—becoming less fragmented and/or more robust—through immersion in a creative task. Honing entails iteratively viewing the creative task from a new context, which may restructure the internal conception of it, and this restructuring may be amenable to external expression. This external change may in turn suggest a new context, and so forth recursively, until the task is complete. The view that creative honing can bring about sweeping changes to an individual's second (psychological) level of complex, adaptive structure is consistent with findings that creativity is potentially therapeutic (Barron, 1963; Forgeard, 2013), and that through immersion in a creative task, a more stable image of the world, and one's own relation to it, can emerge (Pelaprat & Cole, 2011). Thus, it is through the interaction and cross-fertilization of knowledge and ideas that conceptual closure is achieved, and psychological entropy kept to an acceptable minimum. Although the phenomenon cross-domain influence in creativity—and by implication, the abstraction and re-expression of abstract forms—may seem obvious to artists, it plays no role in the bulk of psychology and AI research, in which creativity is portrayed as heuristically guided search or selection amongst discrete, well-defined states, guided by domain-specific expertise (e.g., Simonton, 2010; Weisberg, 1995).

## Implications for Cultural Evolution

At first glance it might seem that the basic units of cultural evolution—i.e., the cultural equivalent of the organism in biological evolution—are such things as catchy songs or rituals or tools. However, the above evidence for the cross-fertilization of different domains suggests that the only way to delineate the cultural lineage of a given idea is to look to the creator's entire loosely-integrated web of knowledge and understandings, i.e., the creative process transforms—not just the problem domain—but the worldview as a whole. In this way, the inspirational sources—or 'conceptual parents'—of a sad ballad could include everything from other musicians, to the patter of rain, to the death of a loved one. Thus, creative products don't just serve practical purposes or provide aesthetic pleasure; they provide tangible external markers of the evolutionary states of the worldviews that generated them. This is consistent with the theory that *what evolves through culture is*, not creative outputs like songs or tools, but *worldviews*, with creative outputs as the externally visible 'excrement' of this transformation process (Gabora, 2004, 2008, 2013). In short, research into cross-domain influences has implications for not just how the creative process works but for how culture evolves.

As discussed elsewhere, a worldview not only self-organizes in response to perturbations but it is imperfectly reconstituted and passed down through culture (Gabora, 2013). This is because it is not just self-organizing but *self-regenerating:* people share experiences, ideas, and attitudes with each other, thereby influencing the process by which other worldviews form and transform. Children expose elements of what was originally an adult's worldview to different experiences, different bodily constraints, and thereby forge unique internal models of the relationship between self and world. In short, worldviews transform by interleaving (1) internal interactions amongst their parts, and (2) external interactions with others. It is through these social interactions that novelty accumulates, and culture evolves.

## Limitations and Future Directions

We attribute the large fraction of respondents who did not complete the questionnaire to the anonymous nature of the study, and the fact that they could receive course credit even if they did not complete it. In future studies, it would be helpful to consider ways of further incentivizing participants to complete the questionnaire. Although the open-ended questionnaire was necessary to enable participants to give anything as an inspirational source, it may have been less inviting for those who prefer a structured format. Changing the format from an online questionnaire to an in-person interview may elicit a higher response rate. In addition, the instructions should actively discourage participants from providing motives (e.g., a desire to be creative) instead of inspirational sources, and examples of each should be given so that the distinction is clear. Also, reformatting the questionnaire so that it is always clear which creative influence is associated with which creative output would help clarify the interpretation of the responses.

Another study is planned to examine the hypothesis that scientists make less use of cross-domain influences than artists. Further research might also investigate developmental differences in the ability to employ cross-domain influences. Future research could investigate the factors that predispose individuals to employ cross-domain influences, such as their implicit conception of how the creative process works. Another direction for future research is to investigate the extent to which the application of cross-domain influences is associated with personality traits correlated with creativity, such as norm-doubting, high aspiration levels, tolerance of ambiguity, and openness to experience (Batey & Furnham, 2006; Eysenck, 1993; Helson, 1996; Martindale & Daily, 1996). If so, this would suggest that propensity toward cross-domain influence plays a mediating role between personality and creativity.

## Practical Implications

The finding that the majority of creative outputs were inspired by cross-domain influences has implications for the development of practices that promote creativity in education, the workplace, and personal life. It suggests that creativity may be cultivated by interdisciplinary courses, as well as activities and cultural objects such as murals of ecological food chains or poetry about science that foster connections between different domains.

## Acknowledgements

We acknowledge and are grateful for funding from grant 62R06523 from the Natural Sciences and Engineering Research Council of Canada.